\documentclass[twocolumn]{aastex62}
\usepackage{CJK, amsmath, amssymb}

\definecolor{cacolor}{RGB}{121,0,102}

\def\postsubmit#1{{#1}}

\graphicspath{{./}{figures/}}

\received{September 3, 2021}
\revised{July 28, 2022}
\accepted{July 31, 2022}
\submitjournal{ApJ}

\shorttitle{CROC Radiation Fields}
\shortauthors{Robinson et al.}

\begin{document}
\begin{CJK*}{UTF8}{gkai}

\title{Can Cooling and Heating Functions be Modeled with Homogeneous Radiation Fields?}

\correspondingauthor{David Robinson}
\email{dbrobins@umich.edu}
\author[0000-0002-3751-6145]{David Robinson}
\affiliation{Department of Physics; University of Michigan, Ann Arbor, MI 48109, USA}
\affiliation{Leinweber Center for Theoretical Physics; University of Michigan, Ann Arbor, MI 48109, USA}

\author[0000-0001-8868-0810]{Camille Avestruz}
\affiliation{Department of Physics; University of Michigan, Ann Arbor, MI 48109, USA}
\affiliation{Leinweber Center for Theoretical Physics; University of Michigan, Ann Arbor, MI 48109, USA}

\author{Nickolay Y. Gnedin}
\affiliation{Particle Astrophysics Center; 
Fermi National Accelerator Laboratory;
Batavia, IL 60510, USA}
\affiliation{Kavli Institute for Cosmological Physics;
The University of Chicago;
Chicago, IL 60637, USA}
\affiliation{Department of Astronomy \& Astrophysics; 
The University of Chicago; 
Chicago, IL 60637, USA}

\begin{abstract}

Cooling and heating functions describe how radiative processes impact the thermal state of a gas as a function of its temperature and other physical properties.
In a most general case the functions depend on the detailed distributions of ionic species and on the radiation spectrum. Hence, these functions may vary on a very wide range of spatial and temporal scales.
In this paper, we explore cooling and heating functions between $5\leq z \leq10$ in simulated galaxies from the Cosmic Reionization On Computers (CROC) project. We compare three functions.  First, the actual cooling and heating rates of hydrodynamic cells as a function of cell temperature.  Second, the median cooling and heating functions computed using median interstellar medium (ISM) properties (median ISM).  Last, the median of the cooling and heating functions of all gas cells (instantaneous).
We find that the median ISM and instantaneous approaches to finding a median cooling and heating function give identical results within the spread due to cell-to-cell variation. 
However, the actual cooling (heating) rates experienced by the gas at different temperatures in the simulations do not correspond to either summarized cooling (heating) functions. In other words, the thermodynamics of the gas in the simulations cannot be described by a single set of a cooling plus a heating function with a spatially constant radiation field that could be computed with common tools, such as Cloudy. 
\end{abstract}


\section{Introduction}\label{sec:intro}

The physics of galaxy formation depends on the interplay between gravitational compression and dissipative cooling of the proto-galactic material.  On smaller scales, this same behavior is replicated in star formation \citep[e.g.][]{binney_77, rees_ostriker, silk_77}.  For example, the cooling rate of gas in a galactic disk can affect the accretion rate of new material onto the disk, which in turn affects the star formation rate \citep{Kannan_2014}.  

Cooling and heating functions describe how the energy density of a gas fluid element changes with time due to radiative processes \citep[e.g.][]{tucker_gould, cox_tucker, sutherland_dopita}. Hence, these functions play an important role in galaxy formation.  The form of these functions depend on the ionization states and energy levels of atoms and ions, and have been extensively calculated and tabulated for solar and cosmic abundances with the assumption that the gas is in Collisional Ionization Equilibrium (CIE).  CIE assumes that there is no incident ionizing radiation and the abundances of various ions are set by a balance between collisional ionization and recombination \citep{cox_tucker, sutherland_dopita, tucker_gould}.  Under the assumption of CIE, the cooling and heating functions depend {\it only} on the gas temperature, density, and metallicity.  Without the CIE assumption, the cooling and heating functions can depend on any physical quantity which affects the ionization states and energy levels of any chemical element in the gas.  In particular, one such quantity is the incident radiation field, specified by its specific intensity $J_\nu$. Radiation fields can ionize and change the energy levels of ions, thus affecting which atomic transitions can emit or absorb photons. Previous works indicate that the presence of ionizing radiation fields can significantly modify the cooling and heating functions of gas \citep[e.g.][]{gnedin_hollon, Wiersma_2009, faerman2021exploring, Romero_2021}.

Since radiation fields impact cooling and heating functions, which, in turn, impact galaxy formation, incident radiation fields can affect galaxy formation in a non-trivial way.  Some examples of this include suppressing of cooling, leading to fewer dwarf galaxies forming \citep{efstathiou} or preventing cooling of gas at $T\sim10^{4-4.5}$ K through suppression of cooling via photoionization of neutral hydrogen \citep{Ceverino_2014}.

Proper modeling of cooling and heating functions therefore requires an understanding of the dependence of cooling and heating functions on the incident radiation field $J_\nu$.  For example, \citet{Wiersma_2009} demonstrates that photoionization by extragalactic radiation can modify cooling functions by about an order of magnitude in realistic conditions.  There are various ways to approach modeling this dependence, with varying accuracy.  The most conceptually straightforward approach is to compute the cooling and heating functions using the full radiation field via a radiative transfer code such as Cloudy \citep{cloudy_98}.  However, this is too computationally expensive to be viable in simulations.  A more practical approach is to tabulate results from Cloudy with gas temperature, density, and metallicity assuming a spatially constant but temporally varying extragalactic background \citep[e.g.][]{haardt_madau, Faucher_Gigu_re_2009, Faucher_Gigu_re_2020}.  Both Illustris \citep{Illustris, AREPO} and EAGLE \citep{EAGLE}, two state of the art cosmological hydrodynamic simulations, use such an approach.  While this approach does incorporate some effects of photoionizing radiation on cooling and heating, it does not account for the radiation field from local sources within galaxies,  which is by far the dominant contribution for ISM gas \citep{EAGLE}.  

\citet{gnedin_hollon} describe a straightforward, albeit approximate, way to incorporate local $J_\nu$ contributions by approximating the full radiation spectrum dependence with 
several key photoionization rates. Hence, one can tabulate the values for cooling and heating functions in a reasonably sized grid in temperature, density, metallicity, and photoionization rates. We will describe this approach in further detail in section~\ref{sec:methodology:chfs}.  \postsubmit{Other works approach including the effect of the local radiation field in different ways.  For example, \citet{Ploeckinger_2020} includes an interstellar radiation field with the same shape as that of the Milky Way, scaled by the local gas properties.  Thus, their approximation scheme does \textit{not} include any radiation field parameters (such as photoionization rates).}

In this paper, we examine the cooling and heating functions of simulated galaxies from the Cosmic Reionization on Computers (CROC) project \citep{gnedin14}.  In these simulations, the dependence of the cooling and heating functions on the local radiation field is computed using the approximation scheme of \citet{gnedin_hollon}.  We compare three approaches to finding the median cooling and heating function across halos of similar mass at a given redshift to see whether the forms of these functions vary, and if so, how. In section~\ref{sec:methods}, we discuss the methodology: the simulations we use, how we calculate the cooling and heating functions, and the three different cooling and heating function medians we consider.  In section~\ref{sec:results}, we demonstrate our results comparing the cooling and heating function medians.  Finally, we close with a summary and discussion of our results and potential future work in section~\ref{sec:conc}.  

\section{Methodology}\label{sec:methods}

\subsection{CROC Simulations}

For this work, we use simulations from the Cosmic Reionization on Computers (CROC) project, a program of simulations utilizing the Adaptive Refinement Tree (ART) code \citep{kravtsov99,kravtsov_etal02,rudd_etal08}.  The simulations include many physical processes expected to be necessary to model cosmic reionization self-consistently.  These processes include gravity, gas dynamics, star formation, stellar feedback, the formation of molecular hydrogen, ionizing radiation from stars and other sources, radiative transfer, and metallicity and radiation field-dependent cooling and heating computed using the approximation scheme of \citet{gnedin_hollon}, which excludes cooling and heating effects due to cosmic rays, molecules, and dust.  The ionizing radiation due to stars is the only ionizing radiation source fully calculated self-consistently.  The simulation incorporates other sources in the radiation background as seen by all regions of the simulation, instead of calculating the contribution from those sources locally.  For more details on the CROC simulations, see \citet{gnedin14}. 

For this work, we use one $20h^{-1}$ comoving Mpc box size simulation realization (denoted box A), which has a spatial resolution of $100$ pc.  CROC resolves the radiation field in both space and time, allowing us to study both the space and time-dependence of cooling and heating functions.

\subsection{Cooling and Heating Functions}\label{sec:methodology:chfs}

We can divide the rate of change of the gas energy density due to radiative processes into two pieces: processes which increase the energy density of the gas (heating processes), and processes which decrease the energy density of the gas (cooling processes).  Guided by this distinction, we can write:

\begin{equation}
    \frac{du}{dt}\bigg|_\text{rad}=n_b^2[\Gamma(T, \ldots)-\Lambda(T, \ldots)],
    \label{eq:CHF_def}
\end{equation}
where $u$ is the energy density of the gas, $n_b$ is the number density of baryons (here, baryons refer to hydrogen, helium, and metal nuclei), and $\Gamma, \Lambda$ are the respective cooling and heating functions of the gas; $T$ is the gas temperature and `$\ldots$' in the cooling and heating functions indicate that these functions generally depend on additional variables besides $T$.  The factor of $n_b^2$ accounts for the baryon number density dependence of collisional processes involving two gas particles.  In collisional ionization equilibrium (CIE), where collisional ionization is balanced by electron recombination, the prefactor of $n_b^2$ ensures that both $\Gamma$ and $\Lambda$ are independent of $n_b$ in the absence of three-body processes \citep{gnedin_hollon}. Note that the left-hand side of equation \ref{eq:CHF_def} only includes changes in energy density due to \textit{radiative} processes, excluding other processes (e.g. gas heating due to adiabatic compression).

While $\Gamma$ and $\Lambda$ are independent of density $n_b$ and depend only on gas temperature and metallicity for gas in CIE, an incident radiation field $J_\nu$ can modify the distribution of energy levels and ionization states of gas particles, changing $\Gamma$ and $\Lambda$.  For an arbitrary $J_\nu$, calculating $\Gamma$ and $\Lambda$ would require a radiative transfer code such as Cloudy \citep{cloudy_98}.  In order to calculate the radiation field-dependent cooling and heating functions, we use the approximation scheme of \citet{gnedin_hollon}.  This is the same approximation that CROC uses to follow gas cooling and heating within the simulation, making this procedure self-consistent \citep{gnedin14}.

This approximation assumes that $J_\nu$ follows a general form including contributions from stars and active galactic nuclei (AGN).  The implementation approximates the $J_\nu$ dependence with 4 key photoionization rates:  HI (neutral hydrogen), HeI (neutral helium), CVI (quintuply-ionized carbon), and the Lyman and Werner bands photodissociation rate for molecular hydrogen (parameterized by $Q_\text{HI}, Q_\text{HeI}, Q_\text{CVI},$ and $Q_\text{LW}$, where $Q_i=P_i/n_b$, and $P_i$ is the photoionization rate for the relevant band). The cooling and heating functions also depend on the local gas temperature $T$, baryon number density $n_b$, and metallicity $Z$. The cooling and heating functions are computed using the radiative transfer code Cloudy for a table of ~4000 values of these parameters, and linear interpolation in each parameter is used for intermediate values \citep{gnedin_hollon}.

This approximation scheme fares well compared to Cloudy calculations within the parameter space of the table described above.  When evaluated on a uniformly log-spaced grid of parameter values (different from the table described above) and compared to the cooling and heating functions calculated by Cloudy, the approximation scheme leads to fractional errors $>2$ in fewer than $1$ in $10^3$ cases. Fractional errors of about $6$ occur in around $1$ in $10^6$ cases \citep{gnedin_hollon}.

\subsection{Cooling and Heating Functions in the Simulations}\label{sec:methodology:inst_flow}

We now consider how to describe the cooling and heating functions in a simulation. To put this question in a broader context, imagine that someone wants to simulate a set of individual galaxies similar to ones modeled in our cosmological simulations. What cooling and heating functions should they adopt?

There are several options of increasing level of complexity:
\begin{description}
\item[A] The simplest option would be to adopt a single set of cooling and heating functions $\mathcal{F}(T)$ (i.e.\ function of temperature only) for the whole simulation. This would be the case if one used, for example, the collisional ionization (CIE) only cooling function \citep{sutherland_dopita}. It is rarely a good approximation, since cooling and heating functions also depend on other gas properties.
\item[B] The next step is to take the cooling and heating functions $\mathcal{F}(T,n,Z)$ that depend on the gas density and metallicity, but ignore the dependence on the spatial variations in the radiation field \citep[such an approximation may still depend on the uniform radiation field, like cosmic background, e.g.][]{kravtsov_etal02,Wiersma_2009}. {\it This is the most commonly used approximation.}  It is the approach used in Illustris-TNG \citep{Illustris, AREPO} and EAGLE \citep{EAGLE}, which both include the effects of a time-varying but spatially homogeneous extragalactic radiation field.
\item[C] Finally, a simulation can adopt the full dependence of the cooling and heating functions $\mathcal{F}(T,n,Z,J_\nu)$ on gas density, metallicity, and the radiation field at every spatial location.  This is the approach used here, and has also been adopted in other simulations such as DRAGONS \citep{DRAGONS3} and Cosmic Dawn \citep{cosmic_dawn}. 
\end{description}

The main question we consider in this paper is how well approximations for cases A and B describe the cooling and heating functions from the case C that is used in our simulations. In order to simplify this question, we consider limited bins in gas density and metallicity within which we can treat densities and metallicities as being approximately constant at the median values $n_0$ and $Z_0$, respectively. Within each bin, cases A and B are identical.

In the simulation, resolution elements (in our case grid cells) within a single bin in density and metallicity have a range of temperatures and radiation fields, and hence also a range of cooling and heating rates $\mathcal{F}_i$ that the cells $i$ have at a given timestep. There are several possible ways one can turn a set of such rates into a single set of cooling and heating functions $\mathcal{F}(T)$.

The most direct way one can define $\mathcal{F}(T)$ is for each value at $T$ to be an average or a median of all cells $i$ that have temperatures $T_i$ sufficiently close to $T$ (say, within a bin of half-width $\Delta T$, i.e. 

\begin{equation}
    \begin{split}
        \bar{\mathcal{F}}_{\rm R}(T) & = \left\langle \mathcal{F}(T_i,n_i,Z_i,J_{\nu, i})\right\rangle_{|T-T_i|<\Delta T} \\
        & \approx \left\langle \mathcal{F}(T_i,n_0,Z_0,J_{\nu, i})\right\rangle_{|T-T_i|<\Delta T} 
    \end{split}
    \label{eq:fchf}
\end{equation}

The second line of the equation emphasizes that, hereafter, we only use cells within single bins of density and metallicity, $|n_0-n_i|<\Delta n$ and $|Z_0-Z_i|<\Delta Z$. We omit dependence on $n_i$ and $Z_i$ for brevity.

Since the cooling and heating rates span several orders of magnitude, we choose the median rather than the average.  We refer to the resulting function from the median rates as the \emph{actual rates}, as it represents the actual cooling and heating rates seen by the cells in the simulation.  
Note that the actual cooling and heating rates $\mathcal{F}_{\rm R}$ do not necessarily correspond to cooling and heating functions for some ``typical" values of density, metallicity, and the radiation field.

The latter can be defined as the cooling and heating functions for the median values of density, metallicity, and the radiation field,

\begin{equation}
    \bar{\mathcal{F}}_{\rm M}(T) = \mathcal{F}(T,n_0,Z_0,\langle J_{\nu, i}\rangle).
    \label{eq:mchf}
\end{equation}

Note that such a median ignores all the information about gas cell temperatures $T_i$, only cell density, metallicity, and the radiation field contribute to that definition. We will refer to this median as the \emph{median ISM} cooling and heating functions.

One can also define another median, which we call ``instantaneous". Imagine a cell $i$ with given values of $T_i$, $n_i$, $Z_i$, and $J_{\nu, i}$. If it heats/cools instantaneously, without any change in $n_i$, $Z_i$, or $J_{\nu, i}$ (which actually change on dynamical and/or star formation timescales), then the temperature of cell $i$ changes with the cooling/heating functions $\mathcal{F}(T,n_i,Z_i,J_{\nu, i})$. Hence, one can define the median instantaneous cooling and heating functions,

\begin{equation}
    \bar{\mathcal{F}}_{\rm I}(T) = \left\langle\mathcal{F}(T,n_0,Z_0,J_{\nu, i})\right\rangle,
    \label{eq:ichf}
\end{equation}
where the median is taken over all cells in the narrow bins of density and metallicity but with any values of temperature or the radiation field.  

The distinction between the three definitions of the cooling and heating functions above is subtle but important. The median ISM cooling and heating functions $\bar{\mathcal{F}}_{\rm M}(T)$ are indeed cooling and heating functions in the canonical sense - they can be computed with, say, Cloudy, for fixed values of $n_0$, $Z_0$, and $J_{\nu, 0}=\langle J_{\nu, i}\rangle$, while the actual cooling and heating rates and the median instantaneous cooling and heating functions may not be.  The physical interpretations of the actual cooling and heating rates and median ISM functions are more clear than for the instantaneous case.

In order to illustrate that a collection of cooling and heating rates is not equal to the cooling and heating function, let us consider a single fluid element with the given density $n_i$, metallicity $Z_i$, and the radiation field $J_i$. Its cooling function is $\Lambda\left(T,n_i,Z_i,J_i(\nu)\right)$. As that fluid element cools thermodynamically (let us assume its heating function is initially small), it may also evolve dynamically.  In general, the density of that element will change (for example, if the cooling is isobaric or adiabatic), unless the cooling is isochoric and no new gas is introduced into the element.  The metallicity and radiation field may also change.  The cooling rate along the flowline of that fluid element is $\dot{E}_C(t) = \Lambda\left(T_i(t),n_i(t),Z_i(t),J_i(\nu,t)\right)$  and the collection of these cooling rates in some time interval, $\{\dot{E}_C\}_{t_0<t<t_1}$ may not correspond to any cooling function $\Lambda(T_i(t),n_*,Z_*,J_{\nu*})$ for some fixed parameters $n_*$, $Z_*$, and $J_{\nu*}$. We show in subsequent sections that this is in fact the case, and highlight the comparison for a range of halo masses and gas metallicities.

\subsection{Median cooling and heating rates and functions of the ISM}\label{sec:methodology:averaging}

We ultimately want to examine how the actual cooling and heating rates and median ISM and instantaneous cooling and heating functions of the interstellar medium of galaxies in our cosmological simulations vary with cosmologically relevant quantities, such as redshift $z$ and host halo mass.  Before we can quantify this behavior\, we first need to examine differences in median cooling and heating rates and functions to capture the average behavior of galaxies in a given mass range and at a given redshift.  

In practice, we consider dark matter halos within the cosmological simulation, rather than individual galaxies.  For each halo identified by the \texttt{ROCKSTAR} \citep{Behroozi_2012} halo catalog at the relevant redshift, we select all gas cells within one virial radius $R_\text{vir}$ of the center of each halo.  For the mass of the halo, we use the virial mass $M_\text{vir}$ from the \texttt{ROCKSTAR} halo catalog.  To account for potential effects due to variations in spatial clustering, we also distinguish between central halos and subhalos using the `parent ID' from the \texttt{ROCKSTAR} halo catalog.  The particular simulation from the CROC project we use for this work has redshift snapshots ranging between $z \sim 5$ and $z \sim 10$. Except for when considering the mass dependence of the cooling and heating functions, we choose a fiducial mass range of $M_\text{vir}>10^{10} \, h^{-1} M_\odot$ (for the rest of this paper, we write $M$ for $M_\text{vir}$, as virial masses are the only masses used here).  

For each combination of mass range, redshift, and choice of central or subhalos, we randomly select 50 halos when they are available, and otherwise select all the halos.  To find and utilize the temperature, number density, metallicity, and photoionization rates of cells within each halo, we used the toolkit \texttt{yt} \citep{yt_11}. We combine the cell data from each of the selected halos, then compute the median cooling and heating functions using the procedures described in section~\ref{sec:methodology:inst_flow}\postsubmit{ for specific bins in density and metallicity (see below)}. At the density scales explored in this paper, the hydrodynamic cells have approximately constant mass.  Hence, in physical terms, the median over hydrodynamic cells is effectively mass-weighted.

We also apply cell selection criteria when calculating the cooling and heating functions.  We examine the cooling and heating functions for cells in the density range of $1<n_b<10$ cm$^{-3}$, the metallicity range $0.03 < Z/Z_\odot < 0.1$ \postsubmit{, where $Z_\odot \approx 0.02$ is the metallicity of gas in the solar neighborhood \citep{gnedin_hollon}}, and with radiation field values $P_\text{HI}, P_\text{HeI} $ $> 0$, and $P_\text{CVI}$ $> 2 \times 10^{-20} \text{s}^{-1}$.  The density range corresponds to the typical galactic ISM.  We limit the metallicity to a relatively narrow range since cooling and heating functions can depend on metallicity. Generally, both cooling and heating functions increase with metallicity across all temperatures, since adding more metals to the gas increases the available pathways for radiative cooling and heating.  Increasing metallicity can also introduce new features (i.e. peaks) to the cooling and heating functions due to elements besides hydrogen and helium. This range \postsubmit{($0.03 < Z/Z_\odot < 0.1$)} ensures that we have sufficient cells for our analysis between redshifts $z \sim 10$ and $z \sim 5$. The photoionization rate cuts are discussed in more detail in the next section.  Unless otherwise noted, all gas cells used in the analysis below are within these parameter ranges.

For each approach described in section~\ref{sec:methodology:inst_flow}, we compute the median cooling and heating rates and functions over all cells remaining after the cuts described above from the selected halos at a given redshift in our fiducial mass bin. As a measure of spread, we compute the 25th and 75th percentiles. For the spread of actual rates and the instantaneous functions, we compute percentiles within all cells in each temperature bin.  For the median ISM cooling and heating functions, we compute the spread by evaluating the cooling and heating functions (in each temperature bin) at the 25th and 75th percentile ISM (that is, the 25th and 75th percentiles of $n_b, Z, P_\text{LW}, P_\text{HI}, P_\text{HeI},$ and $P_\text{CVI}$.  This spread is dominated by spread between cells rather than spread between the medians of different halos. To choose temperature bins, we begin by finding the minimum and maximum cell temperatures for the halos under consideration.  We construct logarithmic temperature bins between these values, with 20 bins per decade.  For ease of comparison, we evaluate the median ISM and instantaneous cooling and heating functions for the same halos at the logarithmic center of the bins used for the actual cooling and heating rates (the geometric mean of the bin edges).  


\subsection{Numerical artifacts}\label{sec:methodology:artifacts}

We exclude cells with any of $P_\text{HI}, P_\text{HeI}, P_\text{CVI}$ equal to $0$ because such cells can yield anomalously large cooling and heating function values with the approximation described in Section~\ref{sec:methodology:chfs}.  The table used to interpolate between photoionization rate values in the approximation does not extend to $P_\text{HI}, P_\text{HeI}, P_\text{CVI}$ this low, so we interpret these cooling and heating function values as numerical artifacts.  Some simulated cells where $P_\text{HI}, P_\text{HeI},$ or $P_\text{CVI}=0$ are obvious numerical artifacts (that is, none of these three photoionization should be 0 for that cell).  It is possible that some cells with zero photoionization rates are \textit{not} numerical artifacts of the simulation, but there is no straightforward way to separate such cells, so we choose to exclude all such cells in this analysis.  This restriction removes a few percent of the cells with density and metallicity in the ranges described in section~\ref{sec:methodology:averaging}.  The fraction of cells removed for the (up to) 50 randomly selected central halos in our fiducial mass bin at various redshifts is shown in Table~\ref{tab:pi_0_frac}.

\begin{table}
\centering
\begin{tabular}{|c|c|}
    \hline
    Redshift $z$ & Fraction of discarded cells ($\%$) \\
    \hline
    5 & 1.5  \\
    8 & 2.1 \\
    9 & 3.2 \\
    10 & 0.7 \\
    \hline
\end{tabular}
\caption{Cells fraction with $P_\text{HI}, P_\text{HeI},$ or $P_\text{CVI}=0$ after density and metallicity cuts.}
\label{tab:pi_0_frac}
\end{table}

\begin{figure}
    \includegraphics[width=\columnwidth]{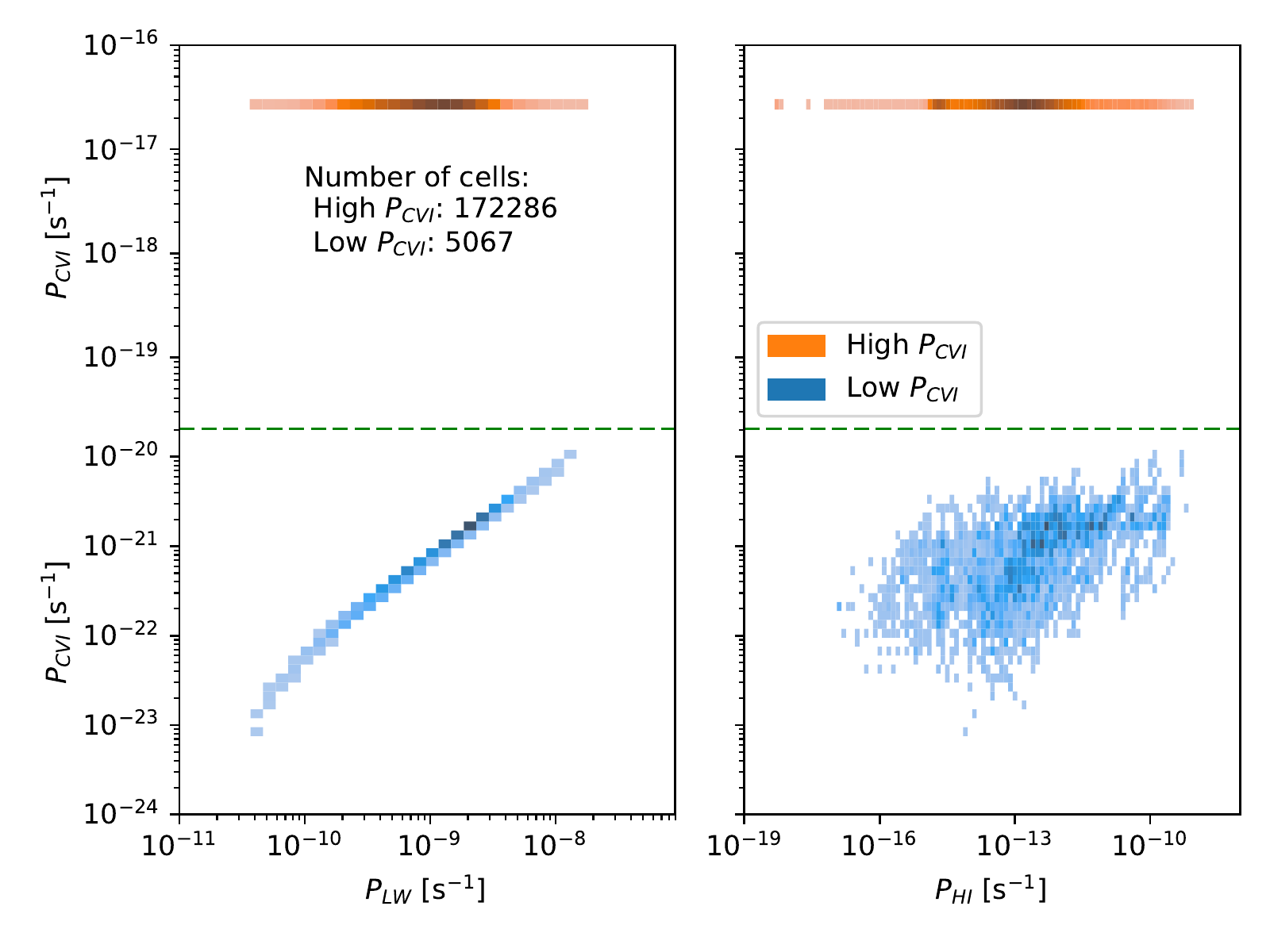}
    \caption{Histograms of the CVI photoionization rate $P_\text{CVI}$ vs the Lyman-Werner band photodissociation rate $P_\text{LW}$ (left) or the HI photoionization rate $P_\text{HI}$ (right) for cells in the 50 randomly selected central halos in the fiducial mass range at $z \sim 5$.  The dashed green horizontal line shows the cutoff between the low and high radiation field parts of the distribution ($P_\text{CVI}=2 \times 10^{-20} \, \text{s}^{-1}$). The two populations of cells are distinguished by color.  The low $P_\text{CVI}$ values and} correlation between $P_\text{CVI}$ and $P_\text{LW}$ for the low $P_\text{CVI}$ population has no obvious physical reason and is likely to be a numerical artifact.
    \label{fig:split_Pcvi_hist}
\end{figure}

Furthermore, we also exclude cells with $P_\text{CVI} < 2 \times 10^{-20}$ s$^{-1}$. We find that the distribution of the CVI photoionization rates, $P_\text{CVI}$, of gas cells in all galaxies in the fiducial mass range is distinctly bimodal across the entire range of simulated redshifts, with no difference in behavior between galaxies hosted by central or subhalos.  A representative plot is shown for our 50 randomly selected central halos at $z \sim 5$ with masses $M > 10^{10} \, h^{-1} M_\odot$ in Figure~\ref{fig:split_Pcvi_hist}.  We term the two distinct regions of the distribution `low $P_\text{CVI}$' for gas populating the distribution in blue and `high $P_\text{CVI}$' for gas populating the distribution in orange.  A cutoff of $2 \times 10^{-20} \, \text{s}^{-1}$ (as shown in the mid-plot horizontal axis in Figure~\ref{fig:split_Pcvi_hist}) separates the two modes at all the redshifts we examined.  The range of $P_\text{CVI}$ for the low $P_\text{CVI}$ distribution does not vary strongly with redshift.  However, the location of the narrow range of high $P_\text{CVI}$ values systematically increases with time from $\sim 7 \times 10^{-20} \, \text{s}^{-1}$ at $z \sim 10$ to $\sim 3 \times 10^{-17} \, \text{s}^{-1}$ at $z \sim 5$ as the cosmic X-ray background gradually builds up.  The cells in each mode are not separated in the phase space of gas properties.  That is, cells in both parts of the $P_\text{CVI}$ distribution are similarly distributed in temperature, number density, and metallicity; $P_\text{CVI}$ is the only feature which cleanly separates the two distributions.  The correlation between $P_\text{CVI}$ and $P_\text{LW}$ for the low $P_\text{CVI}$ population seen in Figure~\ref{fig:split_Pcvi_hist} is surprising.  CVI is ionized by X-rays, sourced only by the spatially constant quasar background in the simulation.  On the other hand, the UV radiation in the Lyman-Werner (LW) bands comes from both local stellar sources and a spatially constant UV background \citep{gnedin14}.  Since these rates originate from disparate sources, there is no obvious reason why they should be correlated. We examined some of these cells visually, and while some of them are located in very high density regions and could be optically thick to X-rays, not all such cells are obviously located in these regions. Hence, we also choose to exclude the cells with low $P_\text{CVI}$, since we cannot clearly show that they are not a previously unidentified numerical artifact of the radiative transfer solver.

Note, no more than a few percent of gas cells (for halos in our fiducial mass range at a given redshift) are cut from our cooling and heating function calculation with these radiation field selection criteria (after the density and metallicity selection criteria have already been applied). 

We find that the fraction of cells with low $P_\text{CVI}$ values (which we remove) varies with redshift.  In Table~\ref{tab:low_pcvi_frac}, we show the fraction of total gas cells in (up to) 50 randomly selected central halos after the density, metallicity, and photoionization rate cuts described above with low $P_{\text{CVI}}$ across various redshifts.  The fraction of low-$P_\text{CVI}$ cells generally increases with increasing redshift.

\begin{table}
\centering
\begin{tabular}{|c|c|}
    \hline
    Redshift $z$ & Fraction of discarded low $P_\text{CVI}$ cells ($\%$) \\
    \hline
    5 &  2.9 \\
    8 & 2.8 \\
    9 & 7.9 \\
    10 & 8.8 \\
    \hline
\end{tabular}
\caption{Redshift evolution of low $P_\text{CVI}$ cell fraction}
\label{tab:low_pcvi_frac}
\end{table}

\section{Results}\label{sec:results}
\subsection{Actual rates vs. median functions}\label{sec:results:flowvsinst}

\begin{figure*}
    \includegraphics{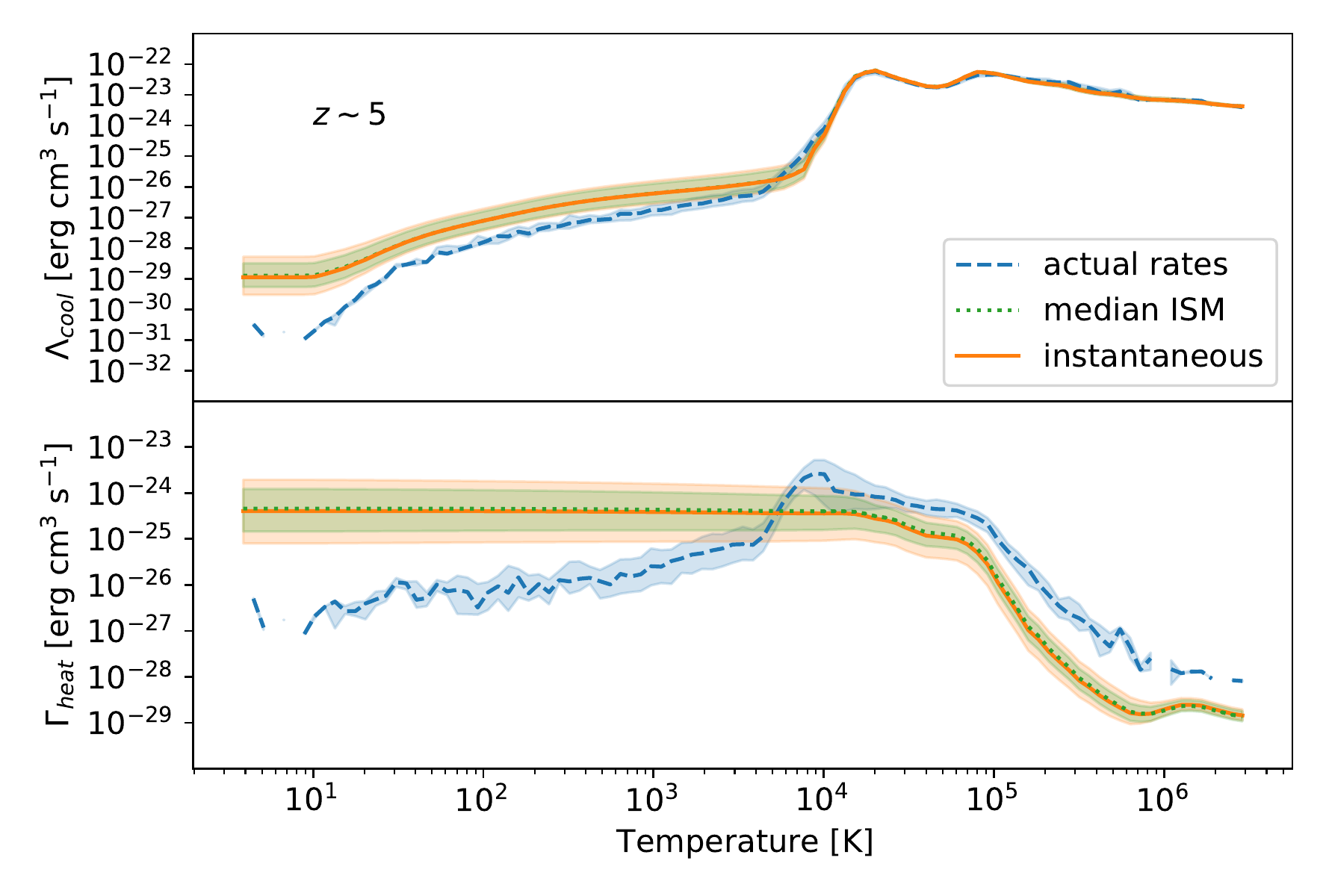}
    \caption{Comparison of actual cooling and heating rates (dashed blue curves) and the median ISM (dotted green) and instantaneous (solid orange) cooling and heating functions for the 50 randomly selected central halos at $z \sim 5$ in the fiducial mass range $M > 10^{10} \, h^{-1} M_\odot$ (the same as shown in Fig.~\ref{fig:split_Pcvi_hist}).  The shaded bands show the 25th-75th percentile spread.} 
\label{fig:fl_vs_inst_comp}
\end{figure*}

We first compare the behavior of the median ISM and instantaneous cooling and heating functions against the actual cooling and heating rates for the 50 randomly selected central halos at $z \sim 5$ in our fiducial mass range, shown in Figure~\ref{fig:fl_vs_inst_comp}.  From Figure~\ref{fig:fl_vs_inst_comp}, we see that the instantaneous and median ISM cooling and heating functions differ from the actual cooling and heating rates by more than the 25th-75th percentile spread below $T \sim 10^3$ K.  They even have different qualitative shapes at these lower temperatures.  Above $T \sim 10^4$ K, there is quantitative agreement for the cooling functions, but only qualitative agreement between the shapes of the heating functions.  On the other hand, the median ISM and instantaneous cooling and heating functions agree very well across all temperature bins, the main difference being that the median ISM cooling and heating functions have slightly smaller spread.  Since the median ISM cooling and heating functions have a clearer theoretical interpretation than the instantaneous case, and these two agree very closely here, we will now focus on the comparison between actual cooling and heating rates against the median ISM cooling and heating functions.

The median ISM cooling and heating functions and actual cooling and heating rates are \textit{not} the same.  This suggests that the actual cooling and heating rates experienced by actual gas cells as their temperature changes are not given by the cooling and heating functions evaluated at median ISM properties.  Hence, we find that the actual cooling and heating rates are \textit{not} expressible in the form $\mathcal{F}(T, n_*, Z_*, J_*(\nu))$ for fixed $n_*, Z_*, J_*(\nu)$ (as explained in section~\ref{sec:methodology:inst_flow}). For example, the actual heating rates do not monotonically decrease with temperature, but all $\Gamma(T, n_*, Z_*, J_*(\nu))$ do. Since the actual cooling and heating rates explicitly describe the thermal evolution of gas cells, we conclude that the thermodynamics of the simulated gas cells cannot be well-described by a single set of cooling and heating functions of the form $\mathcal{F}(T, n_*, Z_*, J_*(\nu))$ , which is a necessary approximation for simulations which use cooling and heating functions with a homogeneous radiation field computed with tools such as Cloudy.  This illustrates how the assumption of a spatially fixed radiation field (the case B approximation discussed in section~\ref{sec:methodology:inst_flow}) can break down in the presence of local radiation fields.  Since we have restricted $1<n_b<10$ cm$^{-3}$ and \postsubmit{$0.03<Z/Z_\odot<0.1$} for these cells, the spread in median ISM cooling and heating functions is primarily due to variations in photoionization rates. We discuss this in further detail below. 

\subsection{Redshift trends}\label{sec:results:redshift}
\begin{figure}
    \includegraphics[width=\columnwidth]{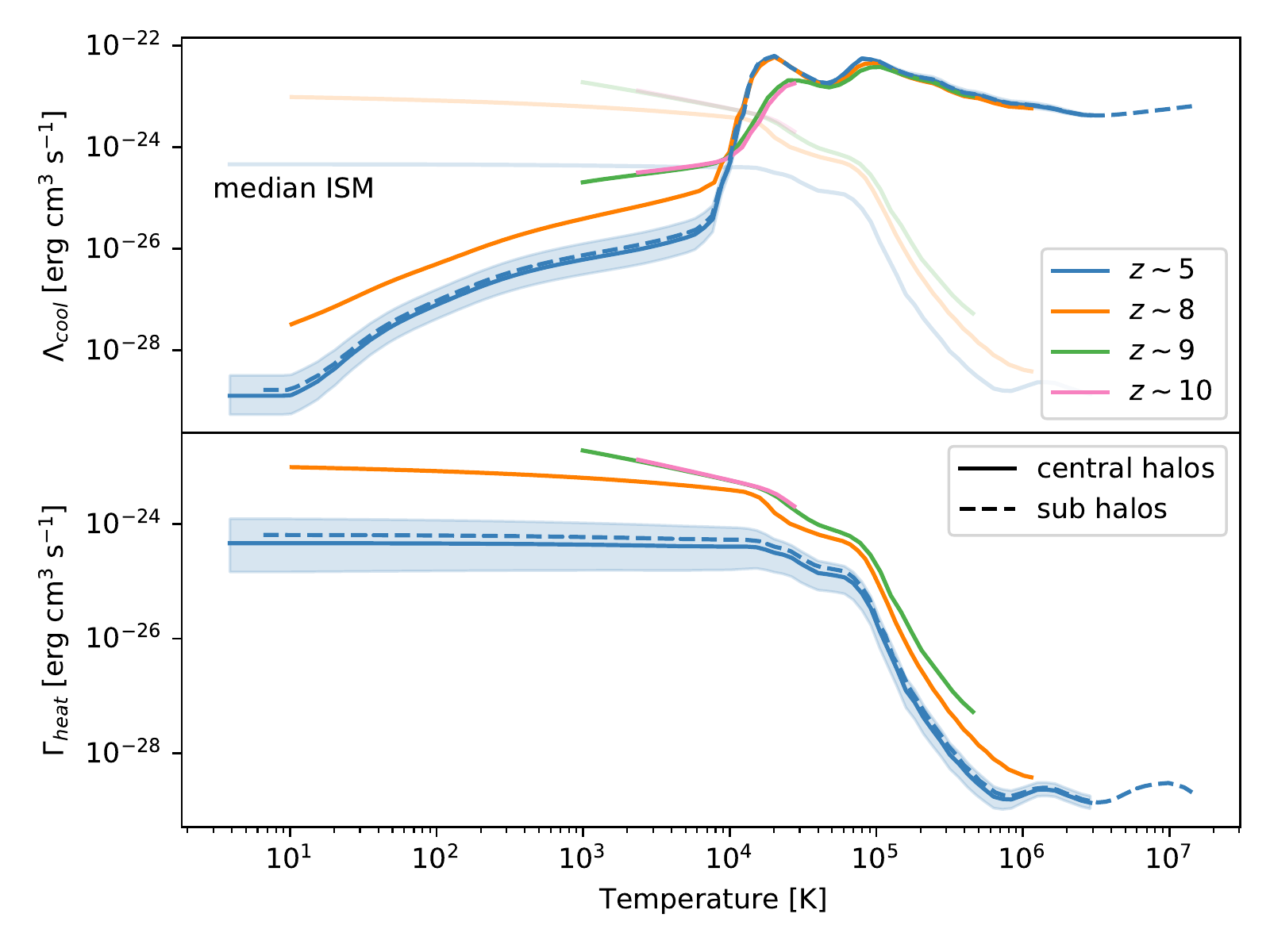}
    \caption{Median ISM cooling and heating functions vs. temperature for up to 50 central halos in the fiducial mass range at $z \sim 5, 8, 9, 10$ (solid lines) and for sub halos at $z \sim 5$ (dashed lines).  The solid lines show the median, while the shaded bands show the 25th-75th percentile spread for central halos at $z \sim 5$.
      For direct comparison, we overplot the heating functions in the upper panel with faint lines; the intersection between the cooling and the heating function corresponds to the equilibrium temperature.}
    \label{fig:z_comp_med_ism}
\end{figure}

The actual cooling and heating rates do not vary strongly with redshift. For simplicity, we do not overplot these.  On the other hand, there is redshift evolution for the median ISM functions.  \postsubmit{Since the number of cells in each temperature bin varies and some temperature bins may contain \textit{no cells at all}, the actual cooling and heating rates are much more `jagged' and less smooth than the median ISM cooling and heating functions (see figures~\ref{fig:fl_vs_inst_comp}, \ref{fig:mass_comp_z5_flowline}, and \ref{fig:metallicity_comp}).  This effect tends to obscure the effects of the redshift evolution seen for the median ISM, discussed below.} Figure~\ref{fig:z_comp_med_ism} shows the median ISM cooling (top panel) and heating (lower panel) functions for gas in central halos in our fiducial mass range at $z \sim 5,8,9,$ and $10$ in the solid curves. Note that the temperature range for each curve is given by the minimum and maximum temperatures of the selected (i.e. within the density, metallicity, and photoionization rate ranges specified in Section~\ref{sec:methodology:averaging}) cells in the up to 50 halos used at the given redshift.  We overplot the heating function in faint lines in the top panel to illustrate the evolution of the intersection point (i.e. equilibrium temperature) between the cooling and heating functions. The equilibrium temperature increases with redshift: from $\sim 10^4$ K at $z \sim 5$ to $2 \times 10^4$ K at $z \sim 10$. For comparison of populations, the dashed lines correspond to the same curves for gas in subhalos at $z \sim 5$, where we have a sufficient number of reasonably resolved subhalos to illustrate a comparison; the central and subhalo populations have indistinguishable median ISM functions within the spread.

From this figure, we see that the median ISM cooling functions increase with redshift for $T \lesssim 10^4$ K and the median ISM heating functions increase with redshift for all temperatures. At $z \sim 9, 10$, the median cooling and heating functions are about three orders of magnitude larger than at lower redshifts, reflecting the weakness of the extragalactic X-ray background in the earliest galaxies before the onset of reionization, reflected in the lower value of $P_\text{CVI}$ (see section~\ref{sec:methodology:artifacts}).  \postsubmit{As the X-ray background increases, it will heat up and ionize gas that was previously cooler and less ionized.  Cooler gas heats more efficiently than hotter gas with the same other properties. Consistent with the expected decrease in cooling efficiency of hotter gas, Figure~\ref{fig:fl_vs_inst_comp} shows that the median ISM heating function is a monotonically decreasing function of temperature.  This overall suggests that the median ISM heating function should decrease as the X-ray background builds up with decreasing redshift, as we see in Figure~\ref{fig:z_comp_med_ism}.} At all the redshifts shown, the cooling function has two peaks (above $\sim 10^4$~K) corresponding to hydrogen and helium.  The location of the first peaks shifts to higher temperatures at $z \sim 9, 10$ compared with $z \sim 5, 8$.

\subsection{Mass dependence}\label{sec:results:mass}

\begin{figure}
    \includegraphics[width=\columnwidth]{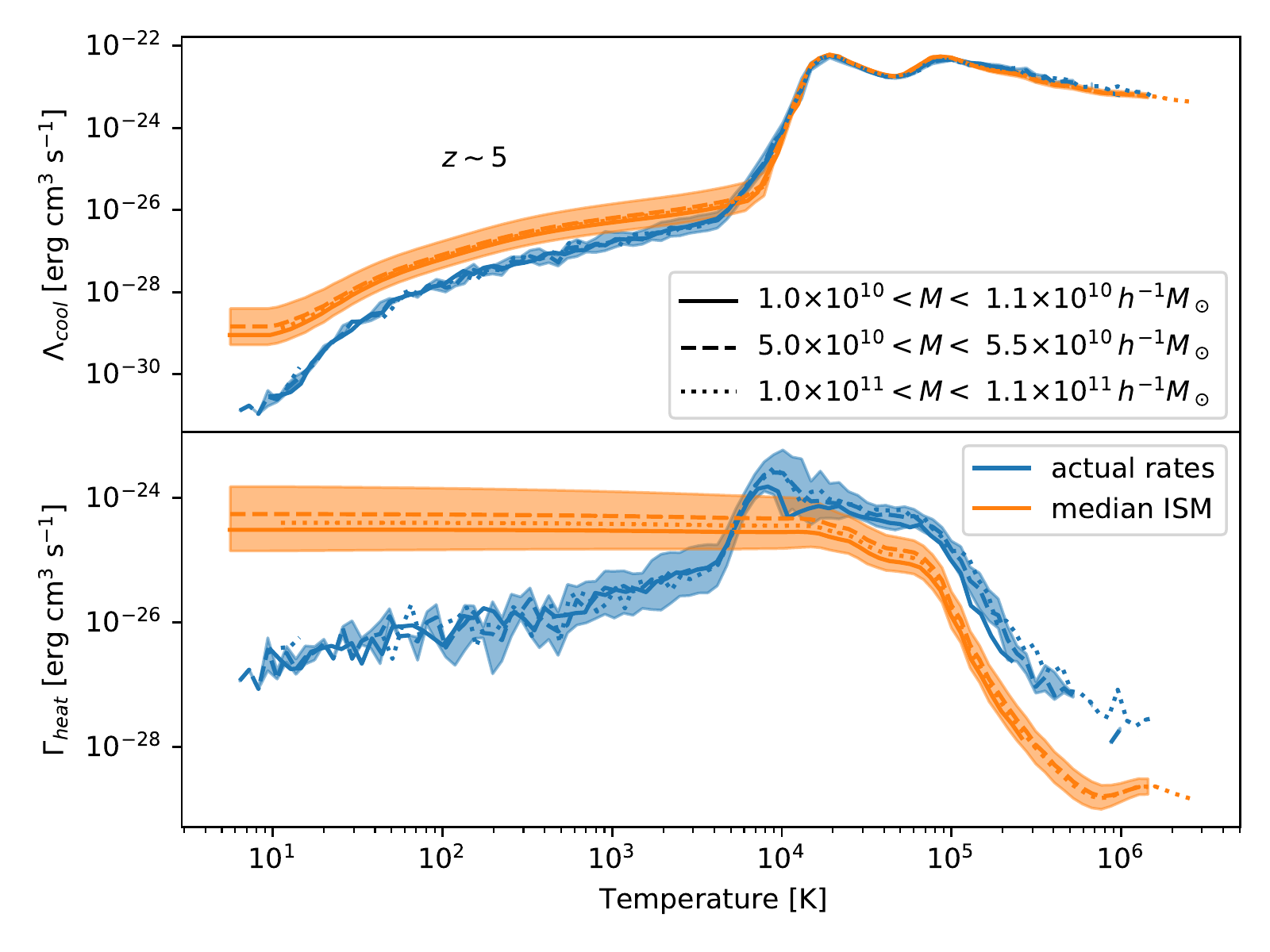}
    \caption{Actual rates (blue) and median ISM functions (orange) for cooling (top panel) and heating (bottom panel) for central halos in three mass bins at $z \sim 5$.  The 25th-75th percentile shaded region is shown only for the middle mass bin, but the widths are similar for all bins.
    }
    \label{fig:mass_comp_z5_flowline}
\end{figure}

Our fiducial halo mass range is relatively wide: $M>10^{10} \, h^{-1} M_\odot$. To justify this choice, we examine the actual cooling and heating rates and median ISM cooling and heating functions in narrow mass bins across the range of halo masses at $z \sim 5$.  We choose mass bins of $(1.0-1.1) \times 10^{10} h^{-1} M_\odot, (5.0-5.5) \times 10^{10} h^{-1} M_\odot $, and $(1.0-1.1) \times 10^{11} h^{-1} M_\odot $. We show the cooling and heating functions in these mass bins at $z \sim 5$ in Figure~\ref{fig:mass_comp_z5_flowline}. 

We observe no halo mass dependence for either the actual cooling and heating rates or median ISM cooling and heating functions at $z \sim 5$.  While there are slight offsets in the curves for the three mass bins at low $T$ ($\lesssim 10^5$ K for heating and $\lesssim 10$ K for cooling), these offsets are much smaller than the 25th-75th percentile bands. Despite the wide choice of fiducial mass bin, neither the actual cooling and heating rates nor the median ISM cooling and heating functions depend on the halo mass within our fiducial mass range.


\subsection{Density and metallicity ranges}\label{sec:nb_Z}
To consider our choices of $1<n_b<10$ cm$^{-3}$ and \postsubmit{$0.03<Z/Z_\odot<0.1$} for density and metallicity ranges, we explore the median ISM cooling and heating functions and actual cooling and heating rates for other density and metallicity ranges (using the same halos as above).

\begin{figure}
    \centering
    \includegraphics[width=\columnwidth]{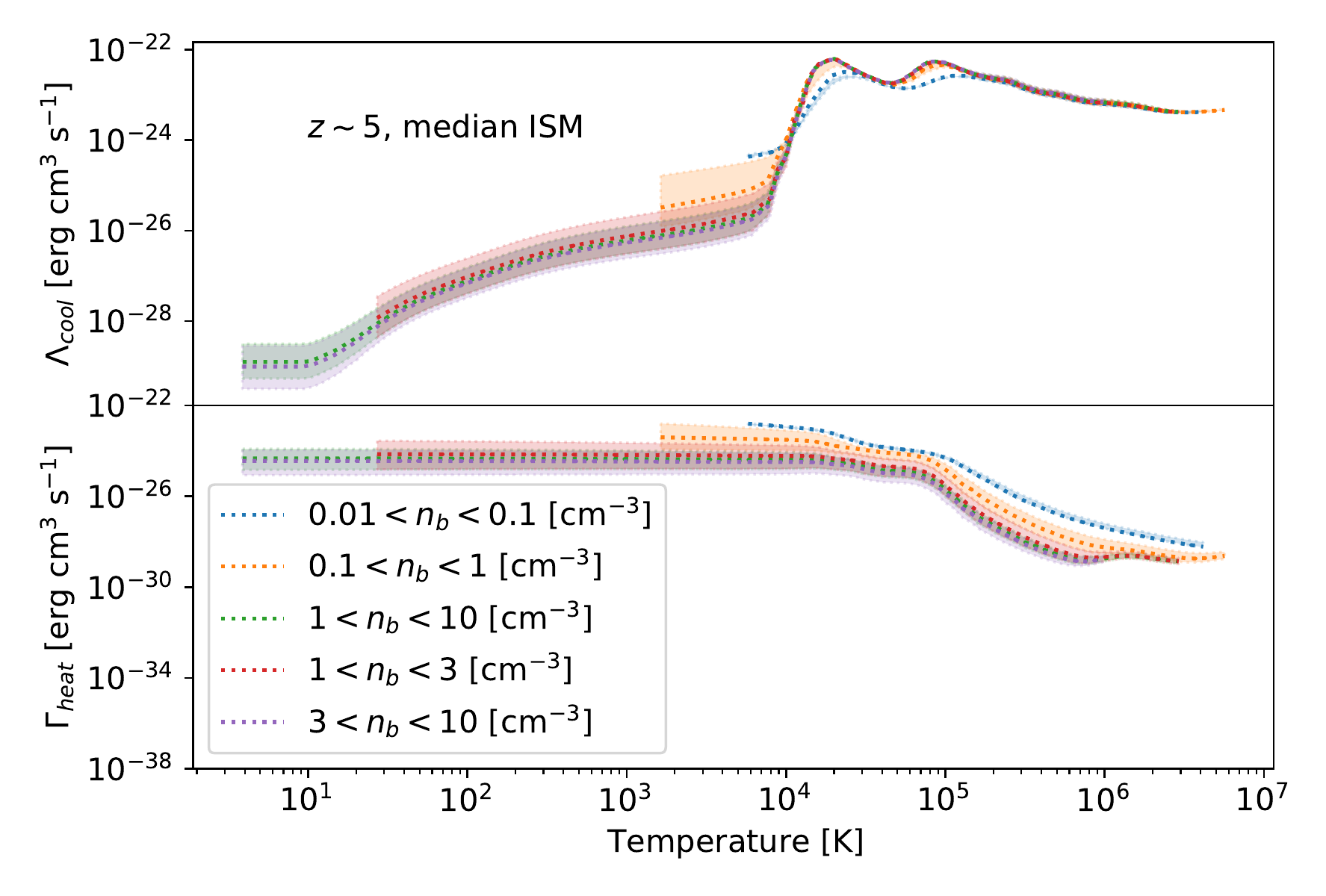}
    \caption{Median ISM cooling and heating functions vs. temperature for selected cells from the random subsample of 50 central halos at $z \sim 5$ in our fiducial mass range (the same sample as previous figures) for cells with $0.01<n_b<0.1, 0.1<n_b<1, 1<n_b<10, 1<n_b<3$ and $3<n_b<10$ cm$^{-3}$. The bands correspond to the 25th to 75th percentile spread for each density range.  
    }
    \label{fig:density_comp}
\end{figure}

First, we consider the density range.  Similar to the lack of redshift dependence, the actual cooling and heating rates also do not vary strongly with the choice of density bin, so we do not overplot these curves.  Instead, we explore how much of the median ISM spread is due to the width of our chosen density range. In Figure~\ref{fig:density_comp}, we plot the median ISM cooling and heating functions for density ranges of $1<n_b<3,$ and $3<n_b<10$ cm$^{-3}$, in addition to $1<n_b<10$ cm$^{-3}$.  For comparison, we also include the median ISM cooling and heating functions for density ranges $0.01<n_b<0.1$ and $0.1<n_b<1$ cm$^{-3}$, which are of the same logarithmic width as our $1<n_b<10$ cm$^{-3}$.  These low density ranges explore the density range of the circumgalactic medium (CGM) rather than the ISM.

As shown in Figure~\ref{fig:density_comp}, the 25th to 75th percentile spreads in the median ISM cooling and heating functions for $1<n_b<3$ and $3<n_b<10$ cm$^{-3}$ have comparable width to that for $1<n_b<10$ cm$^{-3}$, and all three ranges overlap very strongly.  This suggests that the spread within our adopted density range of $1<n_b<10$ cm$^{-3}$ is \textit{not} primarily due to the range of cell densities, but rather to the spread in photoionization rates.  However, the median ISM cooling and heating function spreads for $0.01<n_b<0.1, 0.1<n_b<1,$ and $1<n_b<10$ cm$^{-3}$ overlap only weakly, if at all.  For the lowest density range shown with the blue dotted line, $0.01<n_b<0.1$ cm$^{-3}$, the peaks in the cooling function that correspond to hydrogen and helium shift to higher temperatures (but lower cooling function values). The shift to higher temperatures is due to the fact that fewer atomic collisions occur in low gas density regions, e.g. the CGM.  This gas therefore requires higher temperatures to excite a sufficient number of hydrogen atoms for radiative cooling, which is ultimately less efficient than radiative cooling in a higher density medium. 

\begin{figure}
    \centering
    \includegraphics[width = \columnwidth]{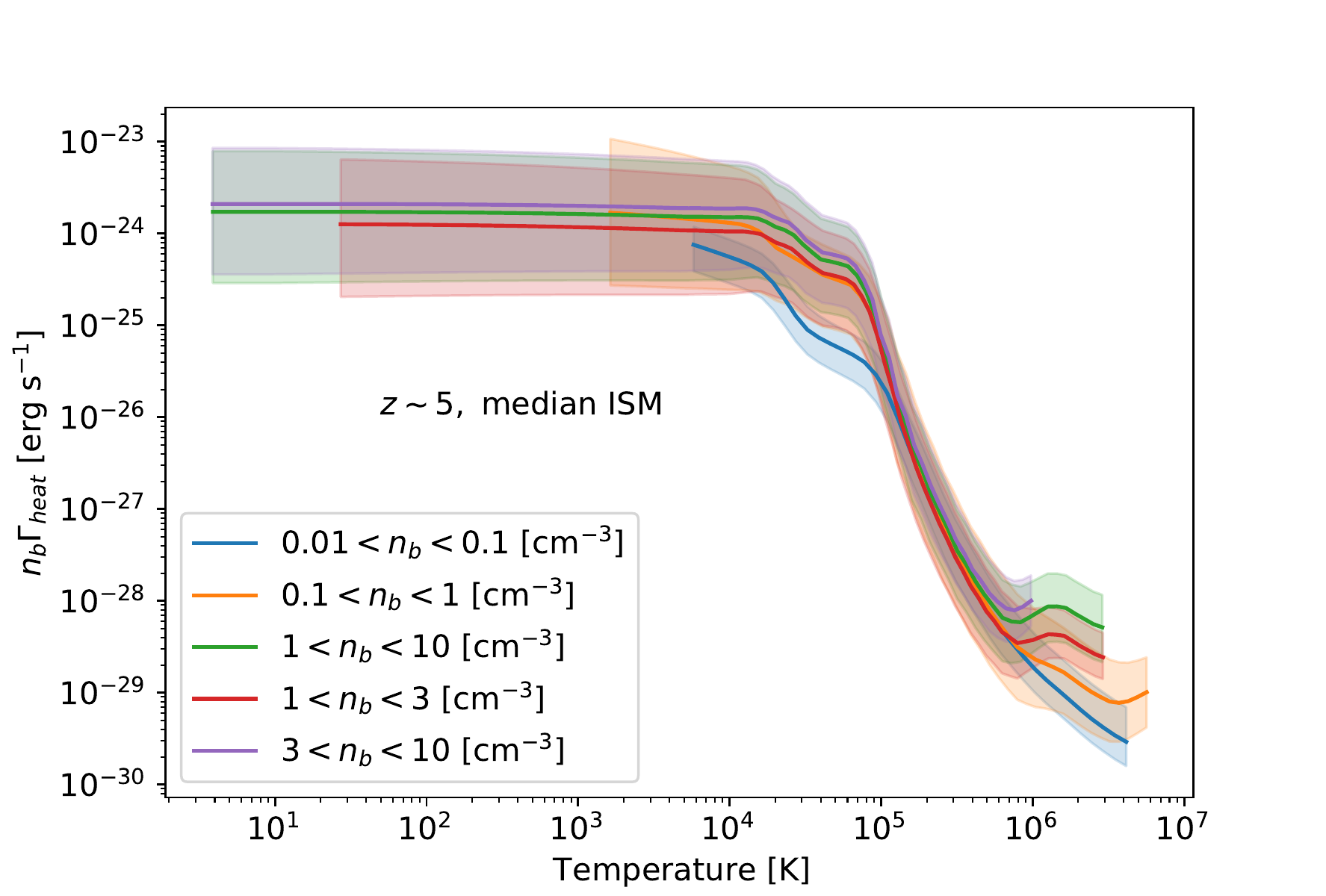}
    \caption{The median baryon number density $n_b$ multiplied by the median ISM heating function vs. temperature for selected cells from the random subsample of 50 central halos at $z \sim 5$ in our fiducial mass range (the same sample as above) for cells with $0.01 < n_b < 0.1, 0.1 < n_b < 1, 1 < n_b < 10, 1 < n_b <3,$ and $3 < n_b < 10 \, \mathrm{cm}^{-3}$.  The shaded regions correspond to the 25th and 75th percentile $n_b$ values respectively multiplied by the heating function evaluated for 25th and 75th percentile ISM properties.}
    \label{fig:n_b_times_HF}
\end{figure}

In general, the heating function includes contributions from direct photoionization heating.  Since this process involves one photon and one atom, the rate of change of energy density is proportional to $n_b$ rather than $n_b^2$ \citep{sutherland_dopita}.  Due to the definition of the heating function $\Gamma(T, \ldots)$ in equation~\ref{eq:CHF_def}, we can expect $\Gamma(T, \ldots) \propto n_b^{-1}$ if photoionization is the only contribution to the heating function.  Thus, for a density range of $1<n_b<10$ cm$^{-3}$, a one dex spread in the cooling and heating functions (at fixed $T, Z,$ and radiation field, i.e. the median ISM case) can be explained by the width of the density range.  It is therefore also useful to compare the product $n_b \Gamma$ for the density ranges discussed above, since $n_b\Gamma$ would be constant across all density ranges if $\Gamma \propto n_b^{-1}$ from photoheating were the only $n_b$ dependence.   We show this product for the median ISM in Figure~\ref{fig:n_b_times_HF}.  Here, we see that $n_b \Gamma$ is indistinguishable for temperatures below $T \sim 4 \times 10^{5}$ K to within the 25th to 75th percentile spread for all density ranges except $0.01 < n_b < 0.1 \, \mathrm{cm}^{-3}$.  The 25th to 75th percentile spread in $n_b \Gamma$ for all density ranges is considerable (larger than 1 dex at the lowest temperatures).  The scatter in $n_b \Gamma$ is primarily due to the scatter in ISM properties other than $n_b$, i.e. the photoionization rates, since we also select a relatively narrow range for the metallicity, $Z$.  \postsubmit{By definition, the radiation field for the median ISM cooling and heating functions is constant.  The photoionization heating rate is therefore also constant.  Thus, if photoionization is the dominant contribution to the heating function (say, over a given temperature range), then the median ISM heating function will also be constant in that regime.}

\begin{figure}
    \centering
    \includegraphics[width=\columnwidth]{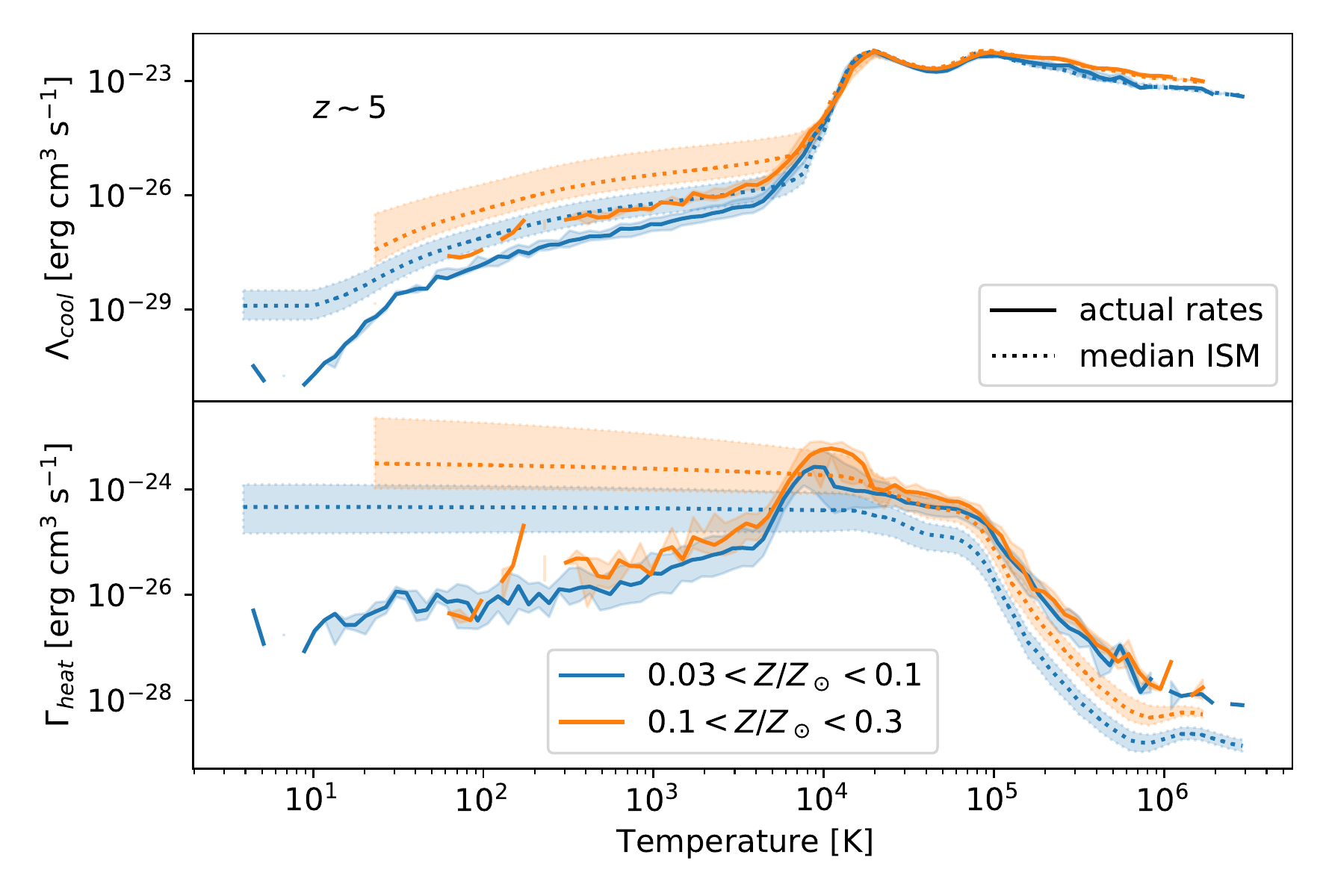}
    \caption{Actual cooling and heating rates and median ISM cooling and heating functions vs. temperature for selected cells from a random subsample of 50 central halos at $z \sim 5$ in our fiducial mass range (the same sample as above) for cells with \postsubmit{$0.03<Z/Z_\odot<0.1$} (as used for the above plots) and \postsubmit{$0.1<Z/Z_\odot<0.3$}.}
    \label{fig:metallicity_comp}
\end{figure}

To consider any potential dependence on our adopted metallicity range of \postsubmit{$0.03<Z/Z_\odot<0.1$},  we examine a different, non-overlapping metallicity range of similar logarithmic width, \postsubmit{$0.1<Z/Z_\odot<0.3$}.  We compare both the actual cooling and heating rates and median ISM cooling and heating functions in Figure~\ref{fig:metallicity_comp}. The qualitative shapes are similar for both metallicity ranges for both actual rates and median ISM functions, but the normalization differs for $T \lesssim 10^4$ K for cooling functions and all temperatures for heating. Our choice of fudicial metallicity range has some effect on the numerical values of the cooling and heating functions at a given redshift, but does not affect our conclusions about the differences between the actual cooling and heating rates and median ISM cooling and heating functions. 

\section{Summary and discussion}\label{sec:conc}

In this work, we compare the actual cooling and heating rates with the median ISM and instantaneous cooling and heating functions of halos in a simulation from the CROC project to assess the validity of universal (i.e. independent of redshift and halo mass) cooling and heating function formulations with a spatially fixed radiation field.  Such formulations are often used in cosmological hydrodynamic simulations, as described in the case B approximation discussed in Section~\ref{sec:methodology:inst_flow}.  Here, cooling and heating functions are computed with tools such as Cloudy using a fixed extragalactic background radiation field. The main conclusions from this work are:

\begin{itemize}
    \item The actual cooling and heating rates cannot be described by the canonical cooling and heating functions of temperature for median values of gas density, metallicity, and the radiation field (see Figure~\ref{fig:fl_vs_inst_comp}). In fact, the heating rates depend non-monotonically on the gas temperature and thus cannot be described by a heating function at \emph{any} fixed values of gas density, metallicity, and the radiation field. Specifically, the commonly used approximation of the gas cooling and heating functions with dependence only on $T, n_b, Z$, and a spatially fixed radiation field is inadequate to describe the actual cooling and heating rates of the ISM illuminated with a spatially varying radiation field (such as with the simulations used here).  While one can still parameterize these rates as functions of temperature, such parameterizations do not have the form of cooling and heating functions which could be computed with with typical, fixed values of the gas density, metallicity, and the radiation field as might be done with codes such as Cloudy.
    
    \item The median ISM cooling and heating functions (a single cooling and heating function evaluated at the median $n_b, Z,$ and photoionization rates) and instantaneous cooling and heating functions (the median of the cooling and heating functions evaluated for the $n_b, Z,$ and photoionization rates of each cell) are nearly identical (see Figure~\ref{fig:fl_vs_inst_comp}).

    \item While there is no significant redshift evolution for the actual cooling and heating rates, the median ISM cooling functions increase with redshift for $T \lesssim 10^4$ K and the median ISM heating function increases with redshift for all temperatures (see Figure~\ref{fig:z_comp_med_ism}), reflecting the weakness of the extragalactic X-ray background from quasars prior to the onset of recombination.  Note, 50\% of the hydrogen becomes ionized between $z \sim 7-8$ in the CROC project simulations considered here \citep{gnedinandkaurov_14}.  
    \item Galaxies hosted by central halos and subhalos exhibit no systematic difference in either their actual cooling and heating rates or their median ISM cooling and heating functions.
    \item There is no significant mass trend in either the actual cooling and heating rates or median ISM cooling and heating functions across a decade of mass within our fiducial mass bin (see Figure~\ref{fig:mass_comp_z5_flowline}).
    \item The scatter in cooling and heating rates is dominated by the scatter within individual halos rather than the scatter between different halos.
\end{itemize}

In our analysis we discovered that a few percent of all cells in CROC simulations have anomalous photoionization rates ($P_\text{HI} = P_\text{HeI} = P_\text{CVI} = 0$ or $P_\text{CVI}$ is suspiciously low), resulting in anomalously high cooling and heating function values in such cells. In future work, we will consider eliminating these numerical artifacts by extending the interpolation tables in the approximation of \citet{gnedin_hollon} to smaller values of $P_\text{HI}$, $P_\text{HeI}$, and $P_\text{CVI}$. In addition to extending the interpolation tables, it may also be possible to improve the approximation (and extend its range of validity further) by using a different combination of photoionization rates in place of the four described in Equation~9 of \citet{gnedin_hollon}.  Machine learning provides a promising tool to assess which combinations of photoionization rates might most impact cooling and heating functions.  A natural follow-up project would be to use these rates to construct new interpolation tables for use in cosmological simulations with radiation fields.

\acknowledgments
This manuscript has been co-authored by Fermi Research Alliance, LLC under Contract No. DE-AC02-07CH11359 with the U.S. Department of Energy, Office of Science, Office of High Energy Physics. This work used resources of the Argonne Leadership Computing Facility, which is a DOE Office of Science User Facility supported under Contract DE-AC02-06CH11357. An award of computer time was provided by the Innovative and Novel Computational Impact on Theory and Experiment (INCITE) program. This research is also part of the Blue Waters sustained-petascale computing project, which is supported by the National Science Foundation (awards OCI-0725070 and ACI-1238993) and the state of Illinois. Blue Waters is a joint effort of the University of Illinois at Urbana-Champaign and its National Center for Supercomputing Applications. This research was also supported in part through computational resources and services provided by Advanced Research Computing (ARC), a division of Information and Technology Services (ITS) at the University of Michigan, Ann Arbor.
CA acknowledges support from the Leinweber Center for Theoretical Physics at the University of Michigan. 
DR acknowledges support from the Physics Department Fellowship at the University of Michigan.
\newpage

\bibliographystyle{apj}
\bibliography{main}

\end{CJK*}
\end{document}